\documentclass[a4paper,fleqn]{cas-dc}
\usepackage{soul}
\usepackage[sort&compress,numbers]{natbib}
\usepackage{lineno}
\usepackage{setspace}
\usepackage{verbatim}
\usepackage{graphics}
\usepackage{graphicx}
\usepackage{cuted}
\usepackage{lipsum}
\usepackage{multirow}
\usepackage[version=4]{mhchem}
\usepackage{chemformula}

\def\tsc#1{\csdef{#1}{\textsc{\lowercase{#1}}\xspace}}
\tsc{WGM}
\tsc{QE}
\tsc{EP}
\tsc{PMS}
\tsc{BEC}
\tsc{DE}

\begin{document}

\let\WriteBookmarks\relax
\def\floatpagepagefraction{1}
\def\textpagefraction{.001}
\shorttitle{CARBON-2D Topological Descriptor}
\shortauthors{Hawthorne \textit{et~al}.}

\title [mode = title]{CARBON-2D Topological Descriptor (C2DTD): An Interpretable and Physics-Informed Representation for Two-Dimensional Carbon Networks}

\author[1,2]{Felipe Hawthorne}
\author[3]{ Marcelo Lopes Pereira Junior}
\author[4]{ Fabiano Manoel de Andrade}
\author[1,2]{ Cristiano Francisco Woellner}
\author[5]{ Raphael Matozo Tromer}

\address[1]{Department of Physics, Federal University of Paraná, 81531-980, Curitiba, PR, Brazil}
\address[2]{Interdisciplinary Center for Science, Technology, and Innovation, Federal University of Paraná, 81531-980, Curitiba, PR, Brazil}
\address[3]{Department of Electrical Engineering, College of Technology, University of Brasília, 70910-900, Brasília, DF, Brazil}
\address[4]{Department of Mathematics and Statistics, State University of Ponta Grossa, 84030-900, Ponta Grossa, PR, Brazil}
\address[5]{Institute of Physics, University of Brasília, 70910-900, Brasília, DF, Brazil}


\begin{abstract}
\noindent Two-dimensional (2D) carbon networks, from pristine graphene to defect-rich and amorphous monolayers, exhibit a complex structure–energy landscape governed not only by local bonding but also by medium-range order and network topology. Capturing these multi-scale effects in a compact, interpretable, and data-efficient manner remains a major challenge for machine learning (ML) in low-dimensional materials. In this work, we introduce the CARBON-2D Topological Descriptor (C2DTD), a physically informed structural representation specifically designed for 2D carbon systems. The descriptor integrates local geometric statistics, a compact radial structural signature, and explicit primitive ring topology into a fixed-length, invariant vector that is both computationally efficient and directly interpretable. Benchmarking on diverse datasets of 2D carbon allotropes and defect-engineered graphene sheets demonstrates that C2DTD achieves robust predictive performance in small-data regimes, outperforming generic high-dimensional featurization schemes while preserving physical transparency. Unsupervised manifold analysis reveals a smoother alignment between descriptor space and the DFT energy landscape, and feature-importance and ablation studies confirm that ring topology emerges as a dominant energetic driver, particularly under vacancy-induced reconstruction. Furthermore, controlled simulations with 5-15\% random vacancies show that C2DTD naturally captures the progressive transition from hexagon-dominated graphene to topologically disordered networks, enabling both dataset-level and structure-specific interpretation. Owing to its compactness, interpretability, and strong physics-based inductive bias, C2DTD provides a fast and generalizable framework for data-driven modeling, defect analysis, and high-throughput screening of 2D carbon materials.
\end{abstract}

\begin{graphicalabstract}
\centering
\includegraphics[width = 0.5\linewidth]{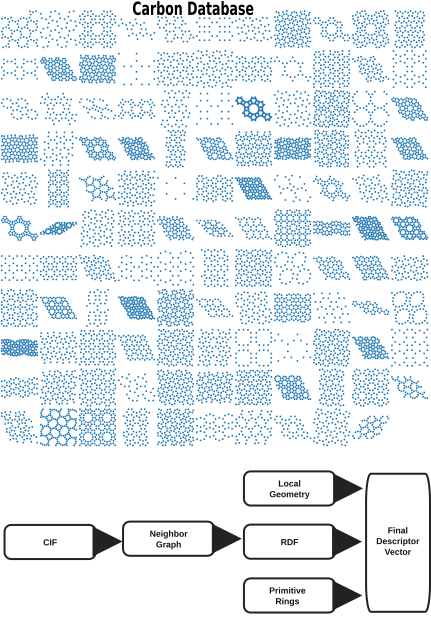}
\end{graphicalabstract}

\begin{highlights}
\item New physics-informed descriptor for 2D carbon networks.
\item Compact representation using local and ring topology.
\item Superior performance in small-data regimes vs matminer.
\item Ring topology emerges as a dominant energetic driver.
\item Interpretable mapping of vacancy-induced reconstructions.
\end{highlights}


\maketitle
\doublespacing

\section*{Introduction}

Two-dimensional (2D) carbon materials, ranging from pristine graphene to defect-engineered sheets, nanoporous networks, and amorphous monolayers, represent one of the most structurally versatile classes of low-dimensional systems \cite{Seol2010, Novoselov2004, Novoselov2007,liu2023defect,najafipour2026multiscale,maroudas2019structure}. Unlike bulk crystals, where periodicity predominantly dictates the structure-property relationship, the physics of 2D carbon networks is intrinsically governed by topology, local coordination, and medium-range reconstruction \cite{yin2024heterogeneous,hwang2026electron}. Small-scale structural perturbations, including vacancies, bond rotations, and the emergence of non-hexagonal rings, induce substantial variations in mechanical stability, electronic transport, and formation energy \cite{Bissett2013,ahmad2021introduction,deshmukh2016nanoscale}. Consequently, the accurate and interpretable modeling of these structure-property relationships remains a formidable challenge in data-driven materials science \cite{Cai2020, Oganov2019, Zhang2017,jiang2025interpretable}.

Machine learning (ML) has emerged as a robust framework for accelerating atomistic modeling, although its efficacy depends critically on the choice of structural descriptors \cite{Chen2019, Bishop2013,Tromer2025}. Traditional hand-crafted features, such as radial distribution functions and coordination metrics, provide physical transparency but frequently fail to encompass the multi-scale complexity of disordered or defect-rich systems. Advanced representations, including atom-centered symmetry functions (ACSF), Smooth Overlap of Atomic Positions (SOAP) descriptors, and libraries such as Materials Data Mining (matminer), offer systematic invariances and broad applicability \cite{Ward2018, Kumar2022, Parsaeifard2021, Xin-yuan2022, Mudassir2022, Schütt2019}. Concurrently, graph neural networks and message-passing architectures learn representations directly from atomic graphs, achieving high accuracy at the expense of increased computational complexity, reduced interpretability, and extensive data requirements \cite{Hakkoum2021, Linardatos2020, Carvalho2019}. These limitations are particularly critical in atomistic studies of low-dimensional materials, where datasets are typically constrained in size and physically heterogeneous.

Despite these advancements, a fundamental gap persists in the description of 2D carbon systems. Most extant descriptors are optimized for chemically diverse bulk materials rather than topology-dominated, mono-elemental, $sp^{2}$-bonded networks \cite{li2023global}. Consequently, such representations often overemphasize local continuous environments or rely on high-dimensional radial encodings, thereby diluting physically significant signals \cite{parsaeifard2022manifolds}. The energetics and stability of 2D carbon are primarily dictated by a subtle interplay between local bonding geometry, medium-range structural order, and the topology of the ring network, specifically the balance between hexagonal and non-hexagonal motifs generated by lattice reconstructions \cite{da2013non,el2022exploring}. A descriptor failing to explicitly encode this interplay is inherently suboptimal for these systems.

In this work, we introduce the CARBON-2D Topological Descriptor (C2DTD), a compact, physically interpretable, and topology-aware structural representation tailored for 2D carbon networks. The central novelty of C2DTD lies in the explicit integration of three complementary physical levels into a fixed-length vector: (i) local geometric statistics capturing coordination, bond-length distributions, angular distortions, and local density; (ii) a compact medium-range radial signature encoding structural organization beyond the first-neighbor shell; and (iii) normalized primitive ring fractions that directly quantify the bonding network topology. By construction, the descriptor is invariant to translations, rotations, and atomic permutations, while maintaining computational efficiency and full interpretability at the feature level.

Conceptually, C2DTD departs from generic featurization strategies by embedding a strong physics-informed inductive bias specific to $sp^{2}$ carbon systems. Rather than representing structures using high-dimensional generic features or opaque latent embeddings, the proposed descriptor directly measures physically meaningful quantities linked to stability mechanisms, such as deviations from the ideal 120$^{\circ}$ bonding geometry and coordination fluctuations induced by vacancies. In this framework, C2DTD bridges the gap between classical, physically motivated descriptors and modern ML requirements, providing both predictive power and mechanistic insight.

The computational efficiency of the C2DTD enables direct calculation from atomic coordinates using neighbor lists and graph-based cycle analysis, thereby circumventing the need for expensive atom-centered expansions or deep neural architectures. This efficiency renders C2DTD particularly suitable for high-throughput screening and defect-engineering studies where rapid evaluation is paramount. Furthermore, the descriptor enables multi-scale analysis, supporting the interpretation of entire datasets through statistical summaries of ring distributions or the decomposition of individual structural signatures.

The inherent interpretability of C2DTD ensures that each feature block corresponds to a distinct physical mechanism: coordination and bond-length statistics quantify local environments, angular distributions capture hybridization strain, and ring statistics explicitly measure topological disorder. This transparency is quantitatively exploitable, as feature importance can be mapped directly onto structural motifs-such as pentagons, hexagons, and heptagons-known to govern the stability of defective graphene. Such a level of mechanistic clarity is rarely achievable with high-dimensional generic descriptors or deep learning models.

Designed for robustness in small-data regimes, C2DTD addresses the constraints of atomistic datasets derived from computationally expensive Density Functional Theory (DFT) or Density Functional Tight-Binding (DFTB) simulations. While data-intensive neural representations may overfit or fail to generalize under these conditions, the compact and physically structured nature of C2DTD facilitates stable learning with tree-based models and limited samples. This characteristic is essential for realistic materials discovery workflows where exhaustive sampling is impractical.

We demonstrate that C2DTD provides a unified, physically faithful representation of diverse 2D carbon systems, ranging from polymorphic allotropes to vacancy-defected graphene. Through systematic benchmarks and defect-engineering case studies, we show that the descriptor achieves competitive predictive performance relative to widely used frameworks while offering unprecedented insight into the topology-energy relationship. More broadly, the proposed framework establishes a general paradigm for descriptor design in low-dimensional materials, in which compactness and topological awareness are essential for bridging ML and atomistic materials physics.

\section*{Methodology}

The C2DTD descriptor is formulated to encode the fundamental structural degrees of freedom governing the energetics of covalent 2D carbon systems. The methodology captures local bonding geometry, medium-range structural order, and primitive ring topology through a deterministic workflow that is invariant to atom indexing and compatible with periodic boundary conditions. Input structures provided in the crystallographic information file format are treated as quasi-2D periodic systems, subject to periodic boundary conditions within the basal plane and kept isolated along the out-of-plane axis. 

A structure is mathematically represented by an atomic set $\mathcal{S} = \{ \mathbf{r}_i \}_{i=1}^{N}$ and a corresponding lattice matrix $\mathbf{H} \in \mathbb{R}^{3 \times 3}$. From this representation, a periodic neighbor graph $G = (V, E)$ is constructed based on a specified spatial cutoff radius $r_c$. Within this graph, vertices correspond to individual atoms, and edges are established according to the minimum-image convention governed by the distance criterion
\begin{equation}
d_{ij} = \left\| \mathbf{r}_j + \mathbf{n}_{ij}\mathbf{H} - \mathbf{r}_i \right\| \leq r_c,
\label{eq:distance}
\end{equation}
where $\mathbf{n}_{ij} \in \mathbb{Z}^3$ denotes the lattice offset vector. This graph structure preserves the essential geometric and topological symmetries required for subsequent feature extraction.

The primary component of the descriptor extracts local geometric statistics directly from the defined neighbor graph. The immediate coordination environment is quantified by the coordination number of atom $i$, defined as
\begin{equation}
C_i = |\mathcal{N}(i)|,
\label{eq:coordination}
\end{equation}
with $\mathcal{N}(i)$ representing the set of neighboring atoms situated within the interaction sphere $r_c$. The relative spatial arrangement is captured via neighbor vectors expressed as
\begin{equation}
\mathbf{v}_{ij} = \mathbf{r}_j + \mathbf{n}_{ij}\mathbf{H} - \mathbf{r}_i.
\label{eq:neighbor_vector}
\end{equation}

These geometric vectors enable the explicit computation of both bond lengths
\begin{equation}
\ell_{ij} = \| \mathbf{v}_{ij} \|,
\label{eq:bond_length}
\end{equation}
and localized bond angles
\begin{equation}
\theta_{ijk} = \cos^{-1} \left( \frac{\mathbf{v}_{ij} \cdot \mathbf{v}_{ik}}{\|\mathbf{v}_{ij}\| \, \|\mathbf{v}_{ik}\|} \right),
\label{eq:bond_angle}
\end{equation}
for all pairs $(j,k)$ belonging to the neighborhood $\mathcal{N}(i)$.

To ensure invariance with respect to global system size and atomic permutation, the framework aggregates these localized metrics into global statistical distributions. For any generic structural quantity $x = \{x_1, \dots, x_M\}$, the representation incorporates the mean value
\begin{equation}
\mu_x = \frac{1}{M}\sum_{m=1}^{M} x_m,
\label{eq:mean}
\end{equation}
along with the corresponding standard deviation
\begin{equation}
\sigma_x = \sqrt{\frac{1}{M}\sum_{m=1}^{M} (x_m - \mu_x)^2},
\label{eq:std}
\end{equation}
as well as the lower and upper limits bounded by $x_{\min}$ and $x_{\max}$.

These specific statistical moments are systematically evaluated for the coordination numbers, bond lengths, and bond angles across the entire lattice. Furthermore, the local structural compactness and short-range density variations are assessed through a Gaussian local density function defined by
\begin{equation}
\rho_i = \sum_{j \in \mathcal{N}(i)} \exp\left(-\frac{\ell_{ij}^2}{2\sigma_d^2}\right),
\label{eq:density}
\end{equation}
incorporating a smoothing parameter $\sigma_d$ to modulate the exponential decay of atomic influence.

The secondary module of the framework encodes medium-range order utilizing a discretized radial distribution function. By evaluating all pairwise distances up to a predefined maximum cutoff $r_{\max}$, the structural ordering is mapped onto a normalized histogram where each bin $g_k$ is computed as
\begin{equation}
g_k = \frac{1}{Z} \sum_{i<j} \mathbb{I} \left( d_{ij} \in [r_k, r_{k+1}) \right),
\label{eq:rdf}
\end{equation}
incorporating the indicator function $\mathbb{I}(\cdot)$ and a normalization constant $Z$ ensuring the integral over all evaluated bins equals unity. 

The third structural component extracts the primitive ring topology from the periodic graph representation. To mitigate boundary artifacts during automated cycle detection, the fundamental unit cell is expanded into a supercell configuration. Primitive rings are strictly identified as chordless cycles within this extended network. Whenever an arbitrary edge $(u,v)$ is temporarily excluded from the connected graph, the associated shortest path $\mathcal{P}_{uv}$ is evaluated to construct a closed structural cycle
\begin{equation}
\mathcal{C} = \mathcal{P}_{uv} \cup \{(u,v)\}.
\label{eq:cycle}
\end{equation}

Valid cycles must inherently lack internal chords and fall strictly within a predefined size spectrum. The topological signature is subsequently formalized through a normalized ring fraction for rings comprising $n$ atoms, expressed mathematically as
\begin{equation}
f_n = \frac{N_n}{\sum_{m} N_m},
\label{eq:ring_fraction}
\end{equation}
where $N_n$ denotes the absolute physical count of primitive rings of size $n$.

The comprehensive descriptor vector integrates all hierarchical structural information by concatenating the independent feature blocks into a unified mathematical object
\begin{equation}
\mathbf{D} =
\left[
\mathbf{S}_{\text{coord}},
\mathbf{S}_{\text{density}},
\mathbf{S}_{\text{bond}},
\mathbf{S}_{\text{angle}},
\mathbf{g}_{\text{RDF}},
\mathbf{f}_{\text{rings}}
\right] \in \mathbb{R}^{d}.
\label{eq:descriptor}
\end{equation}

This continuous formulation inherently preserves the required translational, rotational, and permutational invariances. The mapping between the structural descriptor space and the energetic target variables is established using gradient-boosted decision trees implemented via the XGBoost algorithm \cite{chen2015xgboost}. For a designated training dataset composed of $N$ pairs $\{(\mathbf{D}_i, y_i)\}_{i=1}^{N}$, the model optimizes an objective function
\begin{equation}
\mathcal{L} = \sum_{i=1}^{N} \ell \left(y_i, \hat{y}_i \right) + \sum_{t=1}^{T} \Omega(f_t),
\label{eq:loss}
\end{equation}
combining a squared error loss metric $\ell$ to measure predictive deviation with a structural regularization term $\Omega$ designed to prevent overfitting.

The predictive capability and generalization of the trained models are quantitatively evaluated using the coefficient of determination
\begin{equation}
R^2 = 1 - \frac{\sum_i (y_i - \hat{y}_i)^2}{\sum_i (y_i - \bar{y})^2},
\label{eq:r2}
\end{equation}
alongside the root mean squared error
\begin{equation}
\text{RMSE} = \sqrt{\frac{1}{N} \sum_i (y_i - \hat{y}_i)^2},
\label{eq:rmse}
\end{equation}
and the mean absolute error
\begin{equation}
\text{MAE} = \frac{1}{N} \sum_i |y_i - \hat{y}_i|.
\label{eq:mae}
\end{equation}

Model robustness is rigorously verified by assessing these error metrics across multiple independent train-test splits. The overall framework operates under a constrained set of physically meaningful hyperparameters, primarily consisting of spatial cutoffs, radial discretization limits, and topological ring bounds. The compactness of this parameter space facilitates efficient empirical calibration and ensures strict methodological reproducibility.

\section*{Results}

\begin{figure}[pos = htpb]
\centering
\includegraphics[width=\linewidth]{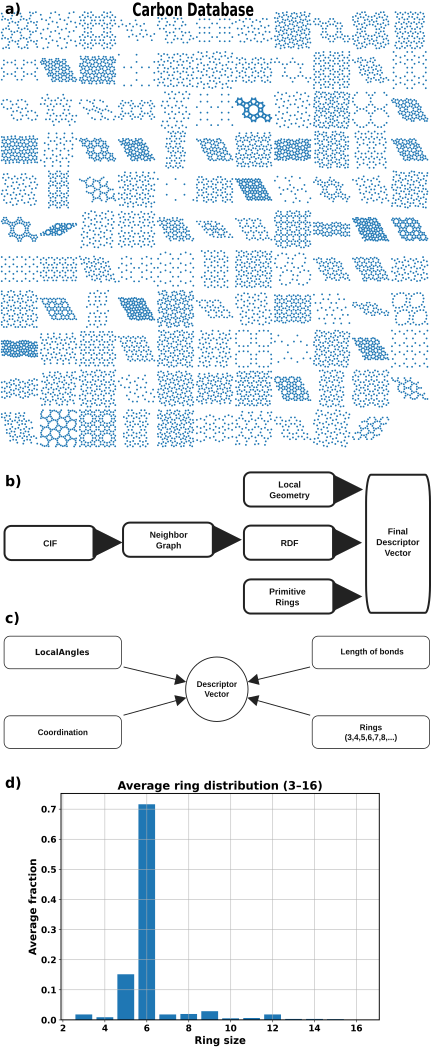}
\caption{Overview of the CARBON-2D Topological Descriptor framework, illustrating (a) the structural diversity of the investigated systems, (b) the descriptor extraction pipeline, (c) the physical mapping of the feature space, and (d) the average ring-size distribution across the dataset.}
\label{fig:fig1}
\end{figure}

The presentation of the results commences with an analysis of the physical foundation and representational architecture of the proposed model. Before evaluating predictive metrics, it is essential to establish how the descriptor captures the dominant energetic degrees of freedom inherent to 2D carbon networks. The primary objective is to construct a compact, physically meaningful, and topology-aware representation. In contrast to generic descriptors designed for chemical universality, the proposed framework is intentionally specialized for covalent $sp^2$ carbon configurations, where thermodynamic stability is strictly governed by local bonding geometry, medium-range structural order, and the topology of the ring network generated by defects and reconstructions. This conceptual foundation is illustrated in Figure~\ref{fig:fig1}.

Figure~\ref{fig:fig1}(a) highlights the structural diversity of the investigated dataset by presenting representative carbon configurations that span a broad range of morphological motifs, from pristine graphene-like regions to highly defect-rich and reconstructed networks. This structural heterogeneity introduces significant challenges for energy prediction models. Multiple configurations share identical chemical compositions and highly similar nearest-neighbor bond lengths, yet exhibit marked differences in total energy. These energetic variations arise from the specific distribution of non-hexagonal rings, defect complexes, and strain fields extending beyond a single coordination shell. In an ML context, this constitutes a highly demanding regime that requires a descriptor capable of resolving subtle yet physically decisive variations in topology and local distortion while rigorously preserving invariance to rigid translations, rotations, and atomic indexing.

The descriptor's conceptual extraction pipeline is detailed in Figure~\ref{fig:fig1}(b). The process begins with the standard crystallographic representation and constructs a periodic neighbor graph using a physically motivated cutoff radius consistent with covalent carbon bonding. The algorithm extracts three complementary information channels directly from this graph and its associated atomic geometry. The initial channel captures local geometry through coordination statistics, bond-length distributions, and angular environments. These specific geometric parameters are strongly correlated with hybridization states and local strain, quantifying deviations from the ideal 120$^\circ$ angle characteristic of $sp^2$ networks. The subsequent channel encodes medium-range order via a compact radial distribution function serving as a coarse spatial fingerprint of structural organization beyond the first coordination shell. The third and most distinctive information channel extracts primitive ring statistics from the periodic graph, providing an explicit topological signature of the network connectivity. All extracted components are concatenated into a final fixed-length vector, enabling direct integration with conventional regression algorithms without the computational burden of atom-centered expansions or opaque latent space embeddings.

Figure~\ref{fig:fig1}(c) emphasizes the inherent physical interpretability built into the descriptor design. Every feature group maps directly to a recognizable structural mechanism. Local coordination metrics reflect bonding environments and defect densities, bond-length statistics capture local atomic relaxations, angular distributions quantify changes in hybridization and strain-induced distortions, and ring statistics explicitly measure the prevalence of distinct connectivity motifs, such as five-, six-, and seven-membered rings. This direct mapping to known physical phenomena is indispensable for analyzing 2D carbon systems, where energetic stability is intricately coupled to the departure from ideal hexagonal connectivity. Developing a descriptor that renders this departure quantitatively measurable guarantees both robust predictive capacity and an interpretable basis for subsequent scientific inference.

To contextualize the data distribution, Figure~\ref{fig:fig1}(d) displays the average ring-size profile of the 120-structure carbon allotrope dataset utilized in this investigation \cite{shi2021high}. The distribution reveals a dominant population of six-membered rings accompanied by smaller yet physically significant contributions from alternative ring sizes. This profile provides an immediate macroscopic structural summary of the dataset and underscores the need for topology-aware features. The overwhelming prevalence of six-membered rings reflects the underlying graphene-like ordering. In contrast, the presence of alternative ring sizes encodes the defect-driven reconstructions that introduce topological strain and alter the global energetic stability. From a data-driven modeling perspective, this statistical distribution indicates that a substantial fraction of the energetic variance can be attributed directly to the magnitude of deviation from hexagonal order and the specific nature of the replacement motifs. The proposed framework makes these topological deviations mathematically explicit, enabling regression models to learn physically consistent stability trends rather than relying on indirect structural proxies.

\begin{figure*}[pos = t!]
\centering
\includegraphics[width=0.7\linewidth]{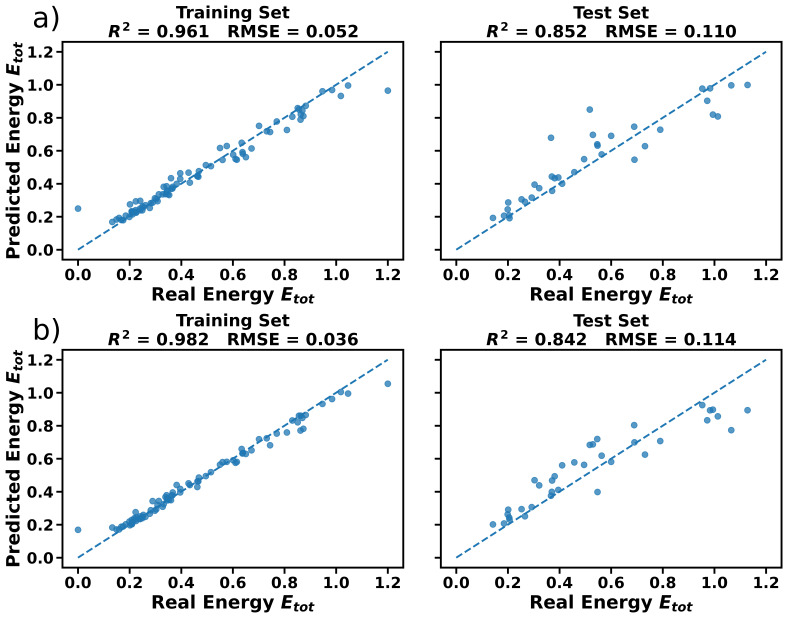}
\caption{Parity plots of predicted versus DFT total energies for 2D carbon structures evaluated at a test size of 0.3, comprising (a) the predictions derived from the C2DTD framework and (b) the predictions obtained utilizing matminer structural features. The dashed lines indicate perfect agreement between predicted and reference energies.}
\label{fig:fig2}
\end{figure*}

To empirically validate this representational alignment, Figure~\ref{fig:fig2} illustrates a representative predictive scenario corresponding to a test fraction of 0.3. This analysis directly compares the predictive performance of the proposed C2DTD framework with that of conventional matminer structural features. Both representations were evaluated under strictly identical ML conditions, utilizing the XGBoost algorithm paired with Optuna hyperparameter optimization. This controlled methodological setup guarantees that any observed performance divergence arises exclusively from the intrinsic representational capacity of the descriptors, rather than from the model architecture, optimization strategy, or data preprocessing.

Within this selected data split, both models achieve high training accuracy. The C2DTD framework yields a training $R^2$ of 0.961, whereas the matminer representation achieves a training $R^2$ of 0.982, indicating that both feature spaces are sufficiently expressive to capture the training data manifold. The decisive comparison emerges in the test regime where generalization capability is rigorously quantified. The C2DTD model achieves a test $R^2$ of 0.852 and a root-mean-square error of 0.110, outperforming the matminer formulation, which yields a test $R^2$ of 0.842 and an error of 0.114. Beyond these numerical metrics, the parity plots reveal a tighter clustering around the ideal prediction line for the topological framework, with reduced dispersion across the entire energy range. This homogeneous error distribution is particularly pronounced in the intermediate-energy regime where structural diversity is highest, suggesting that explicitly encoding local geometry, radial order, and primitive ring topology captures the underlying structure and energy relationship more faithfully than generic representations.

While the single data split provides a clear visual benchmark, a comprehensive assessment extending across multiple training regimes is summarized in Table~\ref{tab:global_benchmark}. The global evaluation demonstrates that the topological descriptor consistently achieves superior predictive performance in eight of nine evaluated test configurations, as measured by both the coefficient of determination and the root mean square error. A marginal exception occurs at a test size of 0.2 where both descriptors exhibit nearly identical predictive accuracy, implying that generic structural features remain competitive under relatively data-rich conditions.

\begin{table}[pos = h]
\centering
\caption{Global comparison of predictive performance metrics between the C2DTD framework and matminer structural features. The table reports the coefficient of determination and the root-mean-square error across varying test-set fractions. The best values for each data split are highlighted in bold.}
\label{tab:global_benchmark}
\begin{tabular}{c|cc|cc}
\hline
\multirow{2}{*}{Test Size} & \multicolumn{2}{c|}{$R^2$} & \multicolumn{2}{c}{RMSE} \\
 & C2DTD & Matminer & C2DTD & Matminer \\
\hline
0.1 & \textbf{0.8886} & 0.8753 & \textbf{0.0859} & 0.0909 \\
0.2 & 0.8912 & \textbf{0.8948} & 0.0937 & \textbf{0.0921} \\
0.3 & \textbf{0.8523} & 0.8424 & \textbf{0.1104} & 0.1141 \\
0.4 & \textbf{0.7357} & 0.6603 & \textbf{0.1494} & 0.1694 \\
0.5 & \textbf{0.7640} & 0.6274 & \textbf{0.1392} & 0.1749 \\
0.6 & \textbf{0.7222} & 0.5881 & \textbf{0.1517} & 0.1847 \\
0.7 & \textbf{0.6605} & 0.5623 & \textbf{0.1599} & 0.1815 \\
0.8 & \textbf{0.5633} & 0.4485 & \textbf{0.1792} & 0.2014 \\
0.9 & \textbf{0.4048} & 0.2129 & \textbf{0.2075} & 0.2386 \\
\hline
\end{tabular}
\end{table}

A fundamental trend emerging from the tabulated global metrics is the systematic divergence in test performance as the effective training set size decreases. In regimes with ample training data, both descriptors achieve comparable accuracy, with modest differences in predictive correlation. As the test size surpasses 0.4, the performance gap becomes increasingly pronounced. At a test fraction of 0.5, the C2DTD framework maintains an $R^2$ of 0.7640, whereas the matminer performance degrades to 0.6274. This systematic divergence amplifies drastically in extremely small-data environments. At a test size of 0.9, the topological descriptor preserves a predictive correlation of 0.4048, effectively doubling the predictive capacity of the matminer representation, which collapses to an $R^2$ of 0.2129.

Because both regression models operate under identical algorithmic constraints, the observed performance divergence is directly attributable to the intrinsic information content of the respective feature spaces. The superior robustness of the C2DTD framework as test sizes increase indicates a stronger inductive bias aligned with the physics of 2D carbon networks. As the volume of available training data decreases, models relying on generic descriptors tend to overfit training-specific correlations and lose predictive stability. This susceptibility is reflected by the sharper degradation of test metrics for the matminer features. The performance of the topological descriptor degrades more smoothly, proving that the formulation captures fundamental thermodynamic stability trends rather than dataset-specific statistical noise.

A critical distinction governing this generalization capability lies in the dimensionality and physical organization of the descriptor spaces. The C2DTD framework is intentionally compact and physically structured, comprising exactly 70 numerical features. This formulation comprises 56 geometric and radial statistics, along with 14 normalized primitive-ring fractions. These components represent aggregated physical observables derived directly from the atomic neighbor graph under 2D periodic boundary conditions. Conversely, the matminer representation is substantially higher-dimensional and dominated by an extended radial distribution function extending up to 20~\AA\ with fine binning. This generic approach generates more than 200 radial features, supplemented by global quantities such as volumetric density and packing fractions. The total dimensionality of the matminer feature space is roughly three times that of the topological descriptor.

This dimensional imbalance directly impacts learning efficiency and explains the widening performance gap in small-data regimes. High-dimensional feature spaces require significantly larger datasets to achieve stable generalization, a phenomenon commonly recognized as the curse of dimensionality. The generic radial vector distributes structural information across hundreds of bins, many of which remain weakly populated or physically irrelevant for $sp^2$ carbon networks. This extensive distribution dilutes the effective signal-to-noise ratio available to the regressor. The matminer model consequently exhibits exceptionally high training accuracy that fails to translate into stable test performance. In contrast, the C2DTD framework executes a physically informed compression of structural data. By employing a compact radial histogram combined with explicit topological descriptors, the framework reduces redundancy while preserving the crucial energetic drivers of 2D carbon.

Beyond predictive robustness and data efficiency, the topological framework offers fundamental advantages in mechanistic interpretability. Each component of the C2DTD has a transparent physical interpretation. Coordination statistics reflect local defect densities, angular distributions quantify hybridization strain, and primitive ring fractions provide a direct measure of topological disorder. This structural transparency enables a direct mapping between model predictions and the underlying physical phenomena. The extended radial representation in generic libraries becomes practically opaque due to high dimensionality and strong inter-feature correlation. Individual radial bins rarely correspond to distinct physical mechanisms, and the inclusion of macroscopic density metrics often captures extrinsic geometric scaling rather than intrinsic bonding physics. The superiority of the topological approach resides primarily in its representational alignment with the governing energetics of 2D carbon allotropes, where stability is definitively controlled by deviations from ideal hexagonal ordering and the resulting topological strain fields.

\begin{figure*}[pos = t!]
\centering
\includegraphics[width=0.7\linewidth]{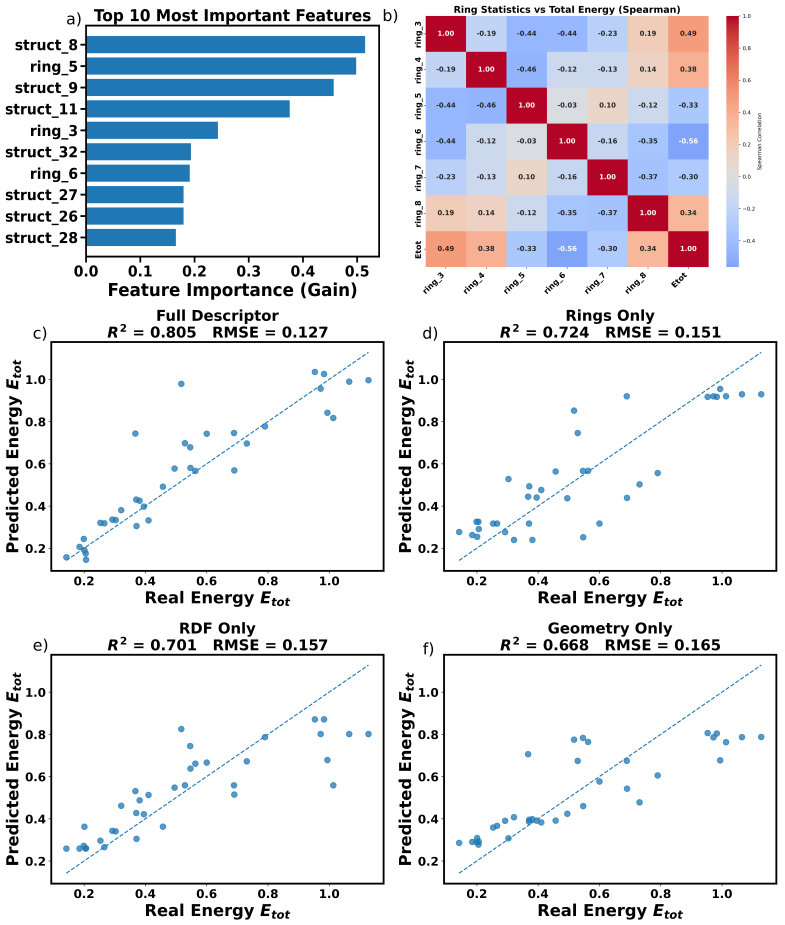}
\caption{Interpretability and ablation analysis of the topological descriptor evaluated at a test fraction of 0.3, depicting (a) the top ten features ranked by relative algorithmic gain, (b) the Spearman rank correlation matrix between normalized ring fractions and total energy, and (c-f) the parity plots and predictive metrics for the full descriptor, the isolated ring statistics, the isolated radial distribution function, and the isolated local geometry features, respectively.}
\label{fig:fig3}
\end{figure*}

To provide a mechanistic rationale for the robust generalization previously established, Figure~\ref{fig:fig3} presents an interpretability and ablation analysis of the proposed framework. Figure~\ref{fig:fig3}(a) reports the top-ranked features determined by the XGBoost gain metric for the identical training configuration utilized in the preceding comparative benchmark. A central outcome is that ring-derived variables emerge among the most influential predictors alongside a restricted subset of compact structural statistics. This result validates the descriptor design by indicating that the regression algorithm relies on physically meaningful quantities rather than diffuse contributions from weakly informative features. The prominence of ring fractions demonstrates that explicit topological information forms a dominant explanatory basis for predicting the total energy in 2D carbon networks.

This topological dependence is further substantiated in Figure~\ref{fig:fig3}(b), which explicitly links ring distributions to energetic stability through Spearman rank correlations between ring fractions and the target total energy. The analysis reveals that rings with five, six, and seven members constitute the only structural motifs exhibiting a negative correlation with total energy. An increased abundance of these specific rings is systematically associated with lower-energy states and, consequently, greater thermodynamic stability. This trend aligns strictly with the established physics of near-ideal hybridization motifs. Six-membered rings correspond to the archetypal hexagonal connectivity of pristine graphene, and their strong negative correlation confirms that maximizing hexagonal order drives the structural manifold toward global energetic minima. The stabilizing contributions of five and seven-membered rings are equally significant, as these specific motifs frequently participate in defect complexes that preserve threefold coordination while facilitating localized structural relaxation and strain redistribution. Conversely, populations of larger or smaller rings exhibit positive correlations, reflecting the physical reality that substantial departures from near-hexagonal topology introduce severe local distortions, degrade medium-range order, and impose significant energetic penalties. The use of Spearman correlation emphasizes monotonic relationships and mitigates sensitivity to outliers, rendering the correlation map a robust quantitative indicator of topological stability trends.

To isolate and quantify the predictive contribution of each fundamental structural block, figures~\ref{fig:fig3}(c-f) present a systematic ablation analysis evaluated under the same testing parameters. The comprehensive descriptor, encompassing all feature groups, achieves the highest accuracy, with an $R^2$ of 0.805 and a root-mean-square error of 0.127. When the model is restricted exclusively to primitive ring statistics, it maintains substantial predictive capacity, yielding an $R^2$ of 0.724 and an error of 0.151. This quantitative retention confirms that network topology alone encodes a dominant fraction of the overarching energy signal, proving that ring distributions are not merely qualitatively correlated with stability but are strictly predictive within a supervised learning framework. The performance of the isolated radial component produces an $R^2$ of 0.701, indicating that medium-range order carries valuable structural information but remains insufficient to fully resolve defect-driven energetic variations without explicit topological context. Similarly, the isolated geometry features yield the lowest baseline performance with an $R^2$ of 0.668, demonstrating that local coordination and angular statistics are physically informative but fundamentally incomplete in isolation.

The systematic performance hierarchy observed across the ablation models establishes a clear representational ordering. Topology provides the strongest independent predictive contribution, while radial and local geometric information supply complementary structural resolution. The superior generalization of the complete framework therefore arises from the synergistic integration of topological motifs that determine the global connectivity class, medium-range radial order that captures organization beyond the first coordination shell, and local geometric metrics that encode specific hybridization and strain effects. This engineered synergy provides a rigorous physics-based rationale for the superior data efficiency and generalization of the topological descriptor relative to generic high-dimensional alternatives.

\begin{figure*}[pos = t!]
\centering
\includegraphics[width=1.0\linewidth]{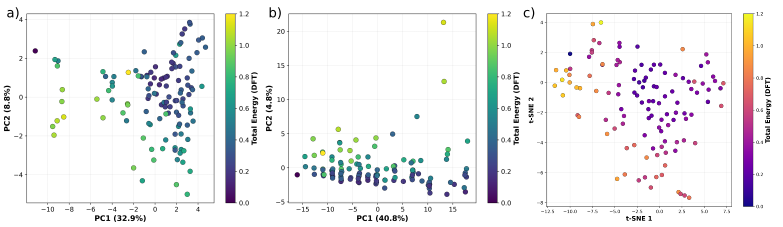}
\caption{Unsupervised dimensionality reduction of the descriptor spaces colored by DFT total energy, showing (a) the principal component analysis projection for the C2DTD framework, (b) the corresponding principal component analysis projection for the matminer structural features, and (c) the non-linear t-distributed stochastic neighbor embedding for the C2DTD framework.}
\label{fig:fig4}
\end{figure*}

To further substantiate the rigorous physics-based rationale derived from the supervised regression analysis, Figure~\ref{fig:fig4} probes the intrinsic geometric organization of the descriptor spaces. This evaluation provides an unsupervised validation complementary to the previous interpretability metrics. All embedded structural representations are colored strictly by their DFT total energy. Under this configuration, the degree of color continuity across the projected manifolds serves as a direct and independent indicator of how smoothly each descriptor parametrizes energetic similarity without reliance on any external regression algorithm.

Figure~\ref{fig:fig4}(a) presents the principal component analysis projection of the topological descriptor space. The primary and secondary principal components capture 32.9\% and 8.8\% of the total variance, respectively, indicating that a substantial fraction of the structural variability is concentrated within a low-dimensional subspace. The embedding exhibits a highly structured energy stratification, with configurations of similar total energy occupying contiguous regions. The resulting color gradient varies smoothly along the dominant projection directions, demonstrating that the framework induces a mathematical representation wherein geometric proximity intrinsically aligns with energetic similarity. This continuous parametrization provides a mechanistic explanation for the superior robustness of the topological descriptor under restricted training data conditions. The observed clustering and gradual energy transitions confirm that the proposed framework successfully isolates the primary structural degrees of freedom governing thermodynamic stability, specifically the degree of hexagonal order and the prevalence of stabilizing topological motifs.

A comparative analysis of the generic matminer descriptor space is provided in Figure~\ref{fig:fig4}(b) through an identical linear projection. Although the primary principal component explains a comparable 40.8\% of the total variance, the secondary component captures only 4.8\%, and the resulting manifold demonstrates a substantially weaker alignment between geometric proximity and energetic similarity. Configurations characterized by drastically different total energies are frequently interspersed within identical regions of the projection plane. This reduced energy stratification indicates that the high-dimensional radial distribution vector and the associated macroscopic density quantities are insufficient to organize 2D carbon structures along energetically meaningful coordinates in an unsupervised setting. The generic representation distributes structural variance across numerous highly correlated radial bins, yielding a feature space in which the directions of maximal mathematical variance do not correspond to the physical directions that govern energy variations in topology-dominated covalent networks. Consequently, the structural mapping becomes disjointed, exacerbating sample inefficiency and driving the sharp degradation in generalization capability observed when the training dataset is reduced.

To capture complex topological relationships that may evade linear projections, Figure~\ref{fig:fig4}(c) visualizes the topological descriptor manifold utilizing t-distributed stochastic neighbor embedding. This non-linear dimensionality reduction algorithm strictly preserves local neighborhood structures within the high-dimensional data. The resulting map reveals highly coherent local groupings characterized by continuous energy gradients, mapping low-energy configurations into compact neighborhoods completely distinct from the regions occupied by highly distorted structures. The emergence of these extended gradients, rather than stochastic intermixing, demonstrates that local neighborhoods within the topological descriptor space correspond strictly to energetically similar physical structures. In 2D carbon networks, this stable neighborhood preservation is vital because energetic states fluctuate strongly in response to topological defects and medium-range connectivity disruptions that are rarely linearly separable. The capacity of the topological descriptor to encode defect topology and geometric distortion into a smooth energy landscape ensures reliable structural interpolation for predictive materials discovery tasks.

\begin{figure*}[pos = t!]
\centering
\includegraphics[width=0.55\linewidth]{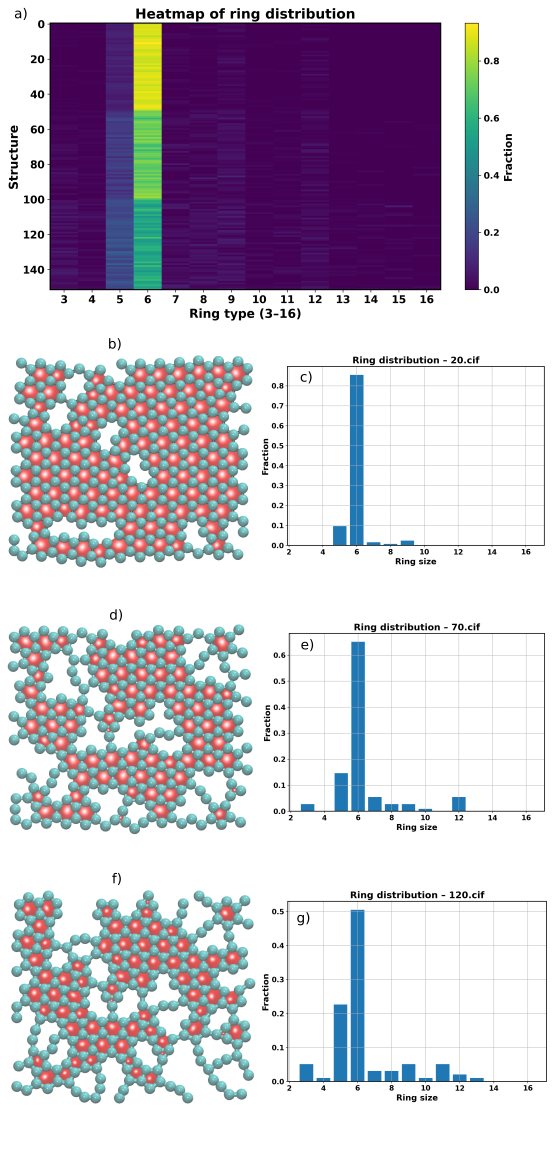}
\caption{Interpretability analysis of the topological descriptor evaluated on a defect-engineered graphene dataset comprising 150 structurally relaxed configurations, illustrating (a) the heatmap of primitive ring fractions across the systematically perturbed dataset, (b-c) the ring distribution for a representative low-defect structure with a 5\% vacancy concentration, (d-e) the corresponding distribution for an intermediate-defect configuration with a 10\% vacancy concentration, and (f-g) the structural profile of a high-defect network containing a 15\% vacancy concentration.}
\label{fig:fig5}
\end{figure*}

Figure~\ref{fig:fig5} provides a physically grounded demonstration of the interpretability of the proposed framework through a controlled defect-engineering dataset. The structural ensemble was constructed from a pristine graphene supercell comprising 288 atoms. One hundred and fifty defective configurations were generated by introducing random vacancy defects and subsequently performing structural relaxation using DFTB methods \cite{Elstner1998,dftb2020}, thereby producing a systematically perturbed distribution of covalent carbon networks. The dataset is partitioned into three distinct defect regimes, with a 5\% vacancy concentration for the first fifty structures, a 10\% concentration for the next fifty, and a 15\% concentration for the final fifty configurations. The spatial distributions of the vacancies were entirely stochastic and explicitly withheld during descriptor construction. This methodology ensures an unbiased assessment of the topological framework's capacity to recover physically meaningful structural transitions, relying exclusively on extracted topology and geometry. From a computational efficiency standpoint, the descriptor evaluation required approximately 10 seconds per structure for systems containing roughly 300 atoms.

Figure~\ref{fig:fig5}(a) presents the heatmap of primitive ring fractions spanning from three-membered to sixteen-membered rings across all evaluated configurations. A highly structured morphological pattern emerges intrinsically, revealing that the descriptor naturally stratifies the dataset according to the underlying vacancy concentration without relying on explicit external supervision. Within the initial 5\% defect regime, the structural profile is overwhelmingly dominated by six-membered rings, reflecting the preservation of the pristine graphene parent lattice. The introduction of sparse vacancies predominantly induces the formation of five-membered rings as a localized reconstruction mechanism designed to minimize dangling bonds while preserving near ideal $sp^2$ coordination. As the defect density increases into the intermediate 10\% regime, the representation shows a systematic reduction in hexagonal motifs, accompanied by an increase in pentagonal content and the gradual emergence of a broader spectrum of ring sizes. This quantitative shift reflects the progressive breakdown of long-range hexagonal order and the onset of topological disorder driven by enhanced lattice reconstruction, bond rotation, and local strain accommodation. 

In the extreme 15\% vacancy regime, the descriptor identifies a pronounced depletion of hexagonal rings and a substantial amplification of alternative polygonal motifs, including rings up to fourteen-membered. This broad distribution constitutes a definitive topological hallmark of severe structural amorphization and the transition from a quasi-crystalline lattice toward a highly disordered covalent network. The framework's capacity to automatically organize these structures into physically distinguishable regimes demonstrates that the descriptor space intrinsically parametrizes defect-induced topological complexity. This dataset-level interpretability confirms that the extracted topological invariants govern the energetic stability and mechanical properties of 2D carbon materials across varying degrees of structural degradation.

To contextualize the global dataset trends at the individual structural level, the lower panels of Figure~\ref{fig:fig5} detail the specific ring distributions for representative configurations across the three defect regimes. Figures~\ref{fig:fig5}(b) and \ref{fig:fig5}(c) examine a low-defect structure characteristic of the 5\% vacancy concentration. The localized ring distribution exhibits a constrained amplitude, dominated almost exclusively by hexagonal motifs, with minor contributions from pentagonal motifs. This tight distribution confirms that the structural backbone remains proximate to the pristine graphene limit where vacancy-induced reconstructions are spatially isolated. Figures~\ref{fig:fig5}(d) and \ref{fig:fig5}(e) illustrate an intermediate-defect case where the underlying ring distribution broadens significantly. The quantitative disparity between hexagonal and pentagonal fractions contracts noticeably, indicating that complex non-hexagonal ring networks begin to aggregate and redistribute local topological strain across the lattice. 

Finally, figures \ref{fig:fig5}(f) and \ref{fig:fig5}(g) detail a highly defective configuration from the 15\% vacancy regime. The framework captures a pronounced compression in the primary ring fractions, alongside a high-amplitude spectrum of polygonal motifs. This pronounced deviation from ideal $sp^2$ ordering provides a rigorous physical quantification of large-scale lattice reconstruction and increased bond-angle variability. The progressive topological evolution captured across these individual panels confirms that the descriptor successfully resolves macroscopic defect regimes while simultaneously detailing the specific local amorphization mechanisms governing individual covalent networks.

\begin{figure*}[pos = t!]
\centering
\includegraphics[width=1.0\linewidth]{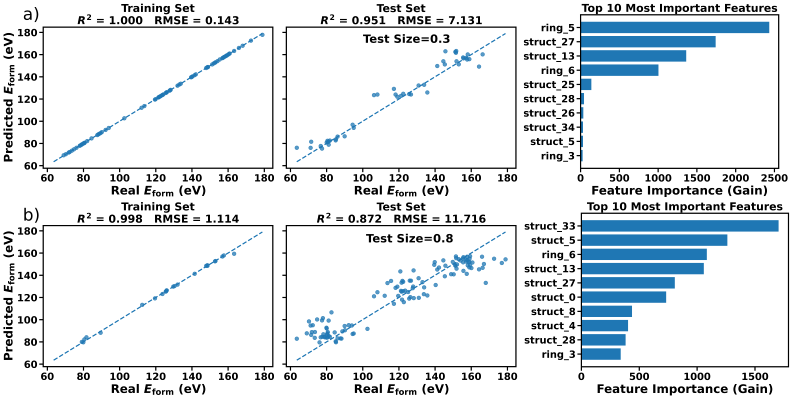}
\caption{Prediction of vacancy-induced formation energy for the defect-engineered graphene dataset utilizing the topological descriptor and gradient-boosted trees, presenting (a) the predictive performance and feature-importance ranking for a test fraction of 0.3 and (b) the corresponding performance and feature attribution under a severe data-scarcity regime with a test fraction of 0.8.}
\label{fig:fig6}
\end{figure*}

Figure~\ref{fig:fig6} extends the interpretability analysis to a predictive setting by modeling the vacancy-induced formation energy of the defect-engineered graphene dataset introduced previously. This ensemble comprises 288-atom sheets with randomly distributed vacancies at concentrations of 5, 10, and 15 percent, which were subsequently relaxed using DFTB methods. The objective of this experiment is to assess whether the topological descriptor can accurately learn the energetics associated with defect introduction and to determine whether the assigned feature importance aligns with physically meaningful topological parameters across varying data-scarcity regimes.

Figure~\ref{fig:fig6} presents the training and test performance for a test size of 0.3 alongside the top ten most important features ranked by algorithmic gain. The model achieves near-perfect training accuracy ($ R^2=1.000$ and error=0.143 eV), while maintaining strong generalization on the test set ($ R^2=0.951$ and error=7.131 eV). These metrics indicate that the framework provides a highly expressive yet physically grounded representation of vacancy-driven energetic variations. The feature importance ranking indicates that ring-based descriptors, particularly five- and six-membered rings, are the dominant contributors to the predictive model. This outcome is fully consistent with the underlying defect physics of graphene. The introduction of vacancies triggers local reconstruction processes that preferentially generate non-hexagonal rings, thereby redistributing strain and modifying the local bonding topology. Consequently, the formation energy becomes strongly coupled to the evolving ring distribution, and the model naturally prioritizes these specific descriptors as the primary predictors.

From a materials physics perspective, the prominence of the five-membered ring fraction as the top feature is especially meaningful. Pentagonal rings represent a well-known structural signature of vacancy reconstruction in $sp^2$ carbon networks because they minimize dangling bonds and reduce local energy penalties through bond rehybridization and lattice relaxation. The simultaneous relevance of the six-membered ring fraction reflects the residual pristine graphene backbone, whose gradual depletion with increasing vacancy concentration directly correlates with the energetic cost of structural disorder. The presence of selected structural statistics, specifically the local geometric distortion and medium-range order components, among the top features further indicates that local geometry complements topological information, reinforcing the multi-scale design philosophy of the descriptor.

Figure~\ref{fig:fig6}(b) presents a more stringent evaluation utilizing a test size of 0.8, corresponding to a severe data-scarcity regime where only twenty percent of the dataset is retained for training. Despite this highly constrained setup, the model achieves a high test performance with an $R^2$ of 0.872 and an error of 11.716 eV, demonstrating remarkable algorithmic robustness and data efficiency. The feature importance profile remains consistent with the lower test-size scenario, in which ring descriptors and a restricted subset of structural features continue to dominate the gain ranking. This stability in feature attribution across drastically different data splits serves as a strong indicator that the model learns intrinsic physical relationships rather than spurious correlations or dataset-specific noise.

The persistence of six-membered, five-membered, and three-membered rings among the most influential features at high test fractions further supports the interpretation that vacancy-induced topological disorder is the primary energetic driver in defective graphene sheets. As the vacancy concentration increases, the progressive reduction of hexagonal rings and the proliferation of alternative polygonal motifs introduce significant local strain, bond rearrangements, and deviations from ideal $sp^2$ hybridization. These fundamental structural transformations directly translate into higher formation energies. The model successfully captures this complex energetic relationship exclusively through topology-aware features rather than relying on global or purely geometric representations.

A critical aspect of these results is the strict consistency between structural interpretability and predictive accuracy. The identical ring statistics that qualitatively explained the defect evolution in the preceding unsupervised analysis emerge quantitatively as the most important predictors in the supervised learning framework. This alignment provides a rare level of transparency in machine learning applied to materials science, allowing algorithmic decisions to be traced directly back to physically interpretable structural motifs. The topological descriptor enables researchers to conclude, in a physically rigorous manner, that the energetic cost of introducing vacancies in graphene is predominantly governed by the redistribution of ring topology and the associated lattice reconstruction.

Overall, the proposed topological framework establishes a robust methodological route for the rapid, interpretable, and data-efficient exploration of 2D carbon materials. This explicit incorporation of primitive ring topology enables future large-scale screening of defective structures, topology-driven design of stable allotropes, and physically guided optimization of nanostructures in which thermodynamic stability, structural disorder, and ring connectivity play a central role.

\section*{Conclusion}

This work introduced the CARBON-2D Topological Descriptor, a compact, physics-informed representation that captures the dominant structural and energetic degrees of freedom in 2D carbon networks. By integrating local geometric statistics, medium-range radial order, and primitive ring topology into a unified framework, the descriptor addresses the inherent limitations of general-purpose featurization strategies, which often fail to capture the topological mechanisms governing $sp^2$-bonded systems. The results demonstrate that C2DTD achieves superior predictive performance and enhanced data efficiency compared to high-dimensional generic descriptors, particularly in the small-data regimes typical of expensive atomistic simulations.

The systematic benchmark against established frameworks proves that the proposed descriptor concentrates physically meaningful information into a structured, low-dimensional vector, ensuring robust generalization as the training set size decreases. This stability arises from a strong inductive bias aligned with the physics of covalent carbon, in which energetic variations are primarily governed by the interplay between lattice distortions and network connectivity. Feature importance and ablation analyses consistently identify ring topology, especially the fractions of five- and six-membered motifs, as the dominant energetic driver. This finding rigorously validates the descriptor design, providing a transparent link between structural motifs and thermodynamic stability that is often obscured in black-box machine learning models.

Unsupervised manifold investigations using PCA and t-SNE further confirm that C2DTD induces a descriptor space that is smoothly aligned with the DFT energy landscape. The natural clustering of energetically similar configurations without explicit supervision underscores the physical faithfulness of the representation. Furthermore, the defect-engineering case study on disordered graphene illustrates the framework's capacity to quantitatively map the transition from hexagon-dominated lattices to topologically amorphous networks, thereby enabling structure-specific interpretation of vacancy-induced reconstructions.

From a computational perspective, the efficiency and scalability of C2DTD render it an ideal candidate for high-throughput screening and the exploration of complex carbon allotropes where interpretability is as critical as predictive accuracy. The integration of topological invariants with geometric and radial descriptors provides a robust foundation for bridging the gap between accurate energy prediction and mechanistic structural insight. Broadly, the design philosophy underlying this work establishes a generalizable framework for developing topology-aware descriptors for other low-dimensional material classes, facilitating the accelerated discovery and rational design of next-generation functional nanomaterials.

\subsection*{Code Availability}

The complete algorithmic implementation of the CARBON-2D Topological Descriptor, encompassing the Python workflow for feature generation and subsequent interpretability analysis, is publicly available in an open-source repository. The source code can be accessed directly at: \\\texttt{https://github.com/tromer-unb/C2DTD-carbon-descriptor}. 

The repository provides the computational scripts to compute the descriptor directly from standard crystallographic information files, along with specialized tools to visualize the induced feature space via primitive ring statistics, topological heatmaps, and dataset-level summaries. A representative minimal dataset is included to guarantee strict methodological reproducibility and facilitate the immediate application of the framework. This open-access distribution ensures transparent validation and promotes the seamless integration of the proposed topology-aware representation into future computational investigations of 2D carbon materials.

\section*{Acknowledgements}

\noindent The authors acknowledge the comprehensive financial support provided by the Coordination for the Improvement of Higher Education Personnel (CAPES), the Brazilian National Council for Scientific and Technological Development (CNPq), the Araucaria Foundation, the Federal District Research Support Foundation (FAPDF), and the São Paulo Research Foundation (FAPESP). Computational resources essential to the execution of this research were generously provided by the National Laboratory for Scientific Computing (LNCC/MCTI, Brazil) through the SDumont supercomputer, as well as the Coaraci Supercomputer and the Center for Computing in Engineering and Sciences at Unicamp. Specifically, FL and CFW received funding from CAPES, CNPq, and the Araucaria Foundation, and acknowledge FAPESP under grants 2019/17874-0 and 2013/08293-7. MLPJ acknowledges support from FAPDF under grant 00193-00001807/2023-16, CNPq under grants 444921/2024-9 and 308222/2025-3, and CAPES under grant 88887.005164/2024-00. FMA was funded by the Araucaria Foundation under project 305 and CNPq under grant 313124/2023-0. RMT acknowledges CNPq for financial support under grants 307371/2025-5 and 444069/2024-0.

\section*{Author Contributions}

\noindent FH: Data curation, Formal analysis, Investigation, Visualization, Writing - original draft. 
MLPJ: Data curation, Formal analysis, Funding acquisition, Investigation, Resources, Validation, Writing - review \& editing. 
FMA: Formal analysis, Investigation, Validation, Writing - original draft.
CFW: Formal analysis, Funding acquisition, Investigation,  Methodology, Resources, Supervision, Writing - original draft.
RMT: Conceptualization, Formal analysis, Data curation, Investigation, Methodology, Project administration, Writing - review \& editing.

\printcredits
\bibliographystyle{unsrt}
\bibliography{bibliography.bib}

@article{Seol2010,
author={Seol, Jae Hun
and Jo, Insun
and Moore, Arden L.
and Lindsay, Lucas
and Aitken, Zachary H.
and Pettes, Michael T.
and Li, Xuesong
and Yao, Zhen
and Huang, Rui
and Broido, David
and Mingo, Natalio
and Ruoff, Rodney S.
and Shi, Li},
title={Two-Dimensional Phonon Transport in Supported Graphene},
journal={Science},
year={2010},
note={ResearchRabbitId e8fae85f-d2df-4ba9-ad61-e670aa87cbc8},
doi={10.1126/SCIENCE.1184014},
url={https://doi.org/10.1126/SCIENCE.1184014}
}

@article{Novoselov2007,
  title={Electronic properties of graphene},
  author={Novoselov, KS and Morozov, SV and Mohinddin, TMG and Ponomarenko, LA and Elias, Daniel Cunha and Yang, Rong and Barbolina, II and Blake, P and Booth, TJ and Jiang, D and others},
  journal={physica status solidi (b)},
  volume={244},
  number={11},
  pages={4106--4111},
  year={2007},
  publisher={Wiley Online Library}
}

@article{Novoselov2004,
  title={Electric field effect in atomically thin carbon films},
  author={Novoselov, Kostya S and Geim, Andre K and Morozov, Sergei V and Jiang, De-eng and Zhang, Yanshui and Dubonos, Sergey V and Grigorieva, Irina V and Firsov, Alexandr A},
  journal={science},
  volume={306},
  number={5696},
  pages={666--669},
  year={2004},
  publisher={American Association for the Advancement of Science}
}

@article{Bissett2013,
author={Bissett, Mark A.
and Konabe, Satoru
and Okada, Susumu
and Tsuji, Masaharu
and Ago, Hiroki},
title={Enhanced Chemical Reactivity of Graphene Induced by Mechanical Strain},
journal={ACS Nano},
year={2013},
month={Nov},
day={26},
publisher={American Chemical Society},
volume={7},
number={11},
pages={10335-10343},
issn={1936-0851},
doi={10.1021/nn404746h},
url={https://doi.org/10.1021/nn404746h}
}

@article{Oganov2019,
author={Oganov, Artem R.
and Pickard, Chris J.
and Zhu, Qiang
and Needs, R. J.},
title={Structure prediction drives materials discovery},
journal={Nature Reviews Materials},
year={2019},
note={ResearchRabbitId 4352db06-0e4b-4282-839b-07272f62237a},
doi={10.1038/S41578-019-0101-8},
url={https://doi.org/10.1038/S41578-019-0101-8}
}

@article{Zhang2017,
author={Zhang, Lijun
and Wang, Yanchao
and Lv, Jian
and Ma, Yanming},
title={Materials discovery at high pressures},
journal={Nature Reviews Materials},
year={2017},
note={ResearchRabbitId 6a53ba22-6d47-4751-88bc-a3c77ebe0ce0},
doi={10.1038/NATREVMATS.2017.5},
url={https://doi.org/10.1038/NATREVMATS.2017.5}
}

@article{Cai2020,
author={Cai, Jiazhen
and Chu, Xuan
and Xu, Kun
and Li, Hongbo
and Wei, Jing},
title={Machine learning-driven new material discovery},
journal={Nanoscale Advances},
year={2020},
note={ResearchRabbitId 19423290-cdf9-4437-bfc3-8e94775c1424},
doi={10.1039/D0NA00388C},
url={https://doi.org/10.1039/D0NA00388C}
}

@article{Chen2019,
author={Chen, Po-Hsuan Cameron
and Liu, Yun
and Peng, Lily},
title={How to develop machine learning models for healthcare.},
journal={Nature Materials},
year={2019},
month={May},
day={01T00:00:00Z},
note={ResearchRabbitId 146692582}
}

@article{Bishop2013,
author={Bishop, Charles M.},
title={Model-based machine learning},
journal={Philosophical Transactions of the Royal Society A: Mathematical, Physical and Engineering Sciences},
year={2013},
month={Feb},
day={13T00:00:00Z},
note={ResearchRabbitId 182419347}
}

@article{Mudassir2022,
author={Mudassir, Mohammed Wasay
and Srinivasan, Sriram Goverapet
and Mynam, Mahesh
and Rai, Beena},
title={Systematic Identification of Atom-Centered Symmetry Functions for the Development of Neural Network Potentials.},
journal={The journal of physical chemistry. A},
year={2022},
note={ResearchRabbitId f9109f19-c2eb-4ed8-85f1-50c178428674},
doi={10.1021/ACS.JPCA.2C04508},
url={https://doi.org/10.1021/ACS.JPCA.2C04508}
}

@article{Schütt2019,
author={Sch{\"u}tt, Kristof T.
and Kessel, Pan
and Gastegger, Michael
and Nicoli, Kim A.
and Tkatchenko, Alexandre
and M{\"u}ller, K.},
title={SchNetPack: A Deep Learning Toolbox For Atomistic Systems.},
journal={Journal of Chemical Theory and Computation},
year={2019},
note={ResearchRabbitId 13ecd332-06c7-4b0e-9915-cb1e8ca6da3a},
doi={10.1021/ACS.JCTC.8B00908},
url={https://doi.org/10.1021/ACS.JCTC.8B00908}
}

@article{Xin-yuan2022,
author={Xin-yuan, Song
and Deng, Chuang},
title={Atomic energy in grain boundaries studied by machine learning},
journal={PHYSICAL REVIEW MATERIALS},
year={2022},
note={ResearchRabbitId de2b1dc6-1f16-4f6d-9786-1805a68ccff0},
doi={10.1103/PHYSREVMATERIALS.6.043601},
url={https://doi.org/10.1103/PHYSREVMATERIALS.6.043601}
}

@article{Parsaeifard2021,
author={Parsaeifard, Behnam
and Goedecker, Stefan},
title={Manifolds of quasi-constant SOAP and ACSF fingerprints and the resulting failure to machine learn four-body interactions.},
journal={Journal of Chemical Physics},
year={2021},
note={ResearchRabbitId dff5586a-8b02-4a99-a612-4342cd981643},
doi={10.1063/5.0070488},
url={https://doi.org/10.1063/5.0070488}
}

@article{Kumar2022,
author={Kumar, Arpan},
title={Virtual Prediction of Material Properties},
journal={Materials Today: Proceedings},
year={2022},
note={ResearchRabbitId dc61d392-999f-42f7-a7dc-8a57df9cba6f},
doi={10.1016/J.MATPR.2022.01.355},
url={https://doi.org/10.1016/J.MATPR.2022.01.355}
}

@article{Ward2018,
author={Ward, Logan
and Dunn, Alexander
and Faghaninia, Alireza
and Zimmermann, Nils
and Bajaj, Saurabh
and Wang, Qi
and Montoya, Joseph H.
and Chen, Jiming
and Bystrom, Kyle
and Dylla, Maxwell
and Chard, Kyle
and Asta, Mark
and Persson, Kristin A.
and Snyder, G. Jeffrey
and Foster, Ian
and Jain, Anubhav},
title={Matminer: An open source toolkit for materials data mining},
journal={Computational materials science},
year={2018},
note={ResearchRabbitId 68a0fbb3-7220-4210-b76c-62d2b7cde39f},
doi={10.1016/J.COMMATSCI.2018.05.018},
url={https://doi.org/10.1016/J.COMMATSCI.2018.05.018}
}

@article{Hakkoum2021,
author={Hakkoum, Hajar
and Abnane, Ibtissam
and Idri, Ali},
title={Interpretability in the medical field: A systematic mapping and review study},
journal={Applied Soft Computing},
year={2021},
note={ResearchRabbitId 89b89cd6-efc4-4a01-a753-33484cc46938},
doi={10.1016/J.ASOC.2021.108391},
url={https://doi.org/10.1016/J.ASOC.2021.108391}
}

@article{Linardatos2020,
author={Linardatos, Pantelis
and Papastefanopoulos, Vasilis
and Kotsiantis, Sotiris},
title={Explainable AI: A Review of Machine Learning Interpretability Methods},
journal={Entropy},
year={2020},
note={ResearchRabbitId bccdecf9-21c9-402f-94a6-9cd61af2343b},
doi={10.3390/E23010018},
url={https://doi.org/10.3390/E23010018}
}

@article{Carvalho2019,
author={Carvalho, Diogo V.
and Pereira, Eduardo M.
and Cardoso, Jaime S.},
title={Machine Learning Interpretability: A Survey on Methods and Metrics},
journal={Electronics},
year={2019},
note={ResearchRabbitId 8d888c37-a5c9-4ada-847d-0d3e3fae094d},
doi={10.3390/ELECTRONICS8080832},
url={https://doi.org/10.3390/ELECTRONICS8080832}
}

@article{Tromer2025,
author={Tromer, Raphael M.},
title={Dynamic Collision Fingerprints (DCF): Introducing a New Descriptor Linking Lattice Interactions to 2D Structural Data Signatures},
journal={Journal of Chemical Theory and Computation},
year={2025},
month={Aug},
day={26},
publisher={American Chemical Society},
volume={21},
number={16},
pages={8106-8118},
issn={1549-9618},
doi={10.1021/acs.jctc.5c00856},
url={https://doi.org/10.1021/acs.jctc.5c00856}
}

@article{liu2023defect,
  title={Defect engineering of two-dimensional materials for advanced energy conversion and storage},
  author={Liu, Fu and Fan, Zhanxi},
  journal={Chemical Society Reviews},
  volume={52},
  number={5},
  pages={1723--1772},
  year={2023},
  publisher={Royal Society of Chemistry}
}

@article{najafipour2026multiscale,
  title={Multiscale modeling of graphene and carbon nanostructures: advances in atomistic, coarse-grained, and machine learning approaches},
  author={Najafipour, Iman and Chaudhuri, Santanu},
  journal={RSC advances},
  volume={16},
  number={11},
  pages={10193--10229},
  year={2026},
  publisher={Royal Society of Chemistry}
}

@article{maroudas2019structure,
  title={Structure-properties relations in graphene derivatives and metamaterials obtained by atomic-scale modeling},
  author={Maroudas, Dimitrios and Muniz, Andre R and Ramasubramaniam, Ashwin},
  journal={Molecular Simulation},
  volume={45},
  number={14-15},
  pages={1173--1202},
  year={2019},
  publisher={Taylor \& Francis}
}

@article{yin2024heterogeneous,
  title={Heterogeneous structured nanomaterials from carbon and related materials},
  author={Yin, Yankun and Hou, Xuyuan and Wu, Bingze and Dong, Jiajun and Yao, Mingguang},
  journal={Advanced Functional Materials},
  volume={34},
  number={49},
  pages={2411472},
  year={2024},
  publisher={Wiley Online Library}
}

@article{hwang2026electron,
  title={Electron Microscopy Approaches to Unraveling the Structure of Amorphous Materials},
  author={Hwang, Sooyeon and Koh, Hyeongjun and Yang, Judith C},
  journal={Small Methods},
  volume={10},
  number={3},
  pages={e01852},
  year={2026},
  publisher={Wiley Online Library}
}

@article{ahmad2021introduction,
  title={Introduction, production, characterization and applications of defects in graphene},
  author={Ahmad, Waqas and Ullah, Zaka and Sonil, Nazmina Imrose and Khan, Karim},
  journal={Journal of Materials Science: Materials in Electronics},
  volume={32},
  number={15},
  pages={19991--20030},
  year={2021},
  publisher={Springer}
}

@article{deshmukh2016nanoscale,
  title={Nanoscale origin and evolution of kinetically induced defects in carbon spheres},
  author={Deshmukh, Sanket A and Narayanan, Badri and Kamath, Ganesh and Pol, Vilas G and Wen, Jianguo and Miller, Dean J and Sankaranarayanan, Subramanian KRS},
  journal={Carbon},
  volume={96},
  pages={647--660},
  year={2016},
  publisher={Elsevier}
}

@article{jiang2025interpretable,
  title={Interpretable machine learning applications: a promising prospect of AI for materials},
  author={Jiang, Xue and Fu, Huadong and Bai, Yang and Jiang, Lei and Zhang, Hongtao and Wang, Weiren and Yun, Peiwen and He, Jingjin and Xue, Dezhen and Lookman, Turab and others},
  journal={Advanced Functional Materials},
  volume={35},
  number={41},
  pages={2507734},
  year={2025},
  publisher={Wiley Online Library}
}

@article{li2023global,
  title={Global mapping of structures and properties of crystal materials},
  author={Li, Qinyang and Dong, Rongzhi and Fu, Nihang and Omee, Sadman Sadeed and Wei, Lai and Hu, Jianjun},
  journal={Journal of Chemical Information and Modeling},
  volume={63},
  number={12},
  pages={3814--3826},
  year={2023},
  publisher={ACS Publications}
}

@article{parsaeifard2022manifolds,
  title={Manifolds of quasi-constant SOAP and ACSF fingerprints and the resulting failure to machine learn four-body interactions},
  author={Parsaeifard, Behnam and Goedecker, Stefan},
  journal={The Journal of Chemical Physics},
  volume={156},
  number={3},
  year={2022},
  publisher={AIP Publishing}
}

@article{da2013non,
  title={Non-hexagonal-ring defects and structures induced by healing and strain in graphene and functionalized graphene},
  author={da Silva-Ara{\'u}jo, Joice and Nascimento, AJM and Chacham, H{\'e}lio and Nunes, RW},
  journal={Nanotechnology},
  volume={24},
  number={3},
  pages={035708},
  year={2013},
  publisher={IOP Publishing}
}

@article{el2022exploring,
  title={Exploring the configurational space of amorphous graphene with machine-learned atomic energies},
  author={El-Machachi, Zakariya and Wilson, Mark and Deringer, Volker L},
  journal={Chemical Science},
  volume={13},
  number={46},
  pages={13720--13731},
  year={2022},
  publisher={Royal Society of Chemistry}
}

@article{shi2021high,
  title={High-throughput screening of two-dimensional planar sp2 carbon space associated with a labeled quotient graph},
  author={Shi, Xizhi and Li, Shifang and Li, Jin and Ouyang, Tao and Zhang, Chunxiao and Tang, Chao and He, Chaoyu and Zhong, Jianxin},
  journal={The Journal of Physical Chemistry Letters},
  volume={12},
  number={47},
  pages={11511--11519},
  year={2021},
  publisher={ACS Publications}
}

@article{chen2015xgboost,
  title={Xgboost: extreme gradient boosting},
  author={Chen, Tianqi and He, Tong and Benesty, Michael and Khotilovich, Vadim and Tang, Yuan and Cho, Hyunsu and Chen, Kailong and Mitchell, Rory and Cano, Ignacio and Zhou, Tianyi and others},
  journal={R package version 0.4-2},
  volume={1},
  number={4},
  pages={1--4},
  year={2015}
}

@article{Elstner1998,
author = {Elstner, M. and Porezag, D. and Jungnickel, G. and Elsner, J. and Haugk, M. and Frauenheim, Th. and Suhai, S. and Seifert, G.},
doi = {10.1103/PhysRevB.58.7260},
issn = {0163-1829},
journal = {Physical Review B},
month = sep,
number = {11},
pages = {7260--7268},
title = {{Self-consistent-charge density-functional tight-binding method for simulations of complex materials properties}},
url = {http://link.aps.org/doi/10.1103/PhysRevB.58.7260},
volume = {58},
year = {1998}
}

@article{dftb2020,
  doi = {10.1063/1.5143190},
  url = {https://doi.org/10.1063/1.5143190},
  year = {2020},
  month = mar,
  publisher = {{AIP} Publishing},
  volume = {152},
  number = {12},
  pages = {124101},
  author = {B. Hourahine and B. Aradi and V. Blum and F. Bonaf{\'{e}} and A. Buccheri and C. Camacho and C. Cevallos and M. Y. Deshaye and T. Dumitric{\u{a}} and A. Dominguez and S. Ehlert and M. Elstner and T. van der Heide and J. Hermann and S. Irle and J. J. Kranz and C. K\"{o}hler and T. Kowalczyk and T. Kuba{\v{r}} and I. S. Lee and V. Lutsker and R. J. Maurer and S. K. Min and I. Mitchell and C. Negre and T. A. Niehaus and A. M. N. Niklasson and A. J. Page and A. Pecchia and G. Penazzi and M. P. Persson and J. {\v{R}}ez{\'{a}}{\v{c}} and C. G. S{\'{a}}nchez and M. Sternberg and M. St\"{o}hr and F. Stuckenberg and A. Tkatchenko and V. W.-z. Yu and T. Frauenheim},
  title = {{DFTB+},  a software package for efficient approximate density functional theory based atomistic simulations},
  journal = {The Journal of Chemical Physics}
}

\end{document}